\documentclass[aps,pre,twocolumn,superscriptaddress]{revtex4-1}
\usepackage{epsfig,amsmath,amssymb,amsfonts,physics,graphicx,color,calc,epstopdf}
\usepackage[ruled,vlined]{algorithm2e}

\usepackage[T1]{fontenc}

\usepackage{xcolor,hyperref}
\hypersetup{
	colorlinks,
	linkcolor={blue!50!black},
	citecolor={blue!50!black},
	urlcolor={blue!80!black}
}

\let\l=\lambda

\def\de{\mathrm{d}}

\newcommand{\beq}{\begin{equation}} 
\newcommand{\eeq}{\end{equation}}
\newcommand{\ba}{\begin{eqnarray}}
\newcommand{\ea}{\end{eqnarray}}

\begin{document}
	
	\title{ Marginal stability of soft anharmonic mean field spin glasses}

	\author{Giampaolo Folena}
	\affiliation{Laboratoire  de  Physique  de  l'Ecole  Normale  Sup\'erieure,  ENS,  Universit\'e  PSL,CNRS,  Sorbonne  Universit\'e,  Universit\'e  de  Paris,  F-75005  Paris,  France}
	\affiliation{ \footnotesize \textit{James Franck Institute and Department of Physics, University of Chicago, Chicago, IL 60637, U.S.A.}}
	
	\author{Pierfrancesco Urbani}
	\affiliation{Universit\'e Paris-Saclay, CNRS, CEA, Institut de physique th\'eorique, 91191, Gif-sur-Yvette, France}

	\begin{abstract}
		We investigate the properties of the glass phase of a recently introduced spin glass model of soft spins subjected to an anharmonic quartic local potential, which serves as a model of low temperature molecular or soft glasses.
		We solve the model using mean field theory and show that, at low temperatures, it is described by full replica symmetry breaking (fullRSB). As a consequence, at zero temperature the glass phase is marginally stable. We show that in this case, marginal stability comes from a combination of both soft linear excitations ---appearing in a gapless spectrum of the Hessian of linear excitations--- and pseudogapped non-linear excitations ---corresponding to nearly degenerate two level systems. Therefore, this model is a natural candidate to describe what happens in soft glasses, where quasi localized soft modes in the density of states appear together with non-linear modes triggering avalanches and conjectured to be essential to describe the universal low temperature anomalies of glasses.
	\end{abstract}

	\maketitle
	
	\paragraph*{Introduction ---}
	One of the biggest open problems in glass physics is the explanation of the anomalous low temperature properties of glasses. Indeed, at low temperatures, these systems are robustly found to be characterized by an abundance of soft modes/low energy excitations with respect to crystalline solids. On the one hand, non-phononic linear excitations are found, which display a universal behavior at low frequencies, with a quartic tail in the density of states (DOS) and quasi localized eigenvectors \cite{modes_prl_2016,MSI17,das2021universal,richard2020universality,bonfanti2020universal,angelani2018probing}. On the other hand, when deformed, structural glasses rearrange in a plastic way, with system spanning avalanches typically described by non-linear excitations, whose density is pseudogapped \cite{oyama2021unified}.  When quantum fluctuations are accounted for, one typically observes an exceptionally high, linear in temperature, specific heat as compared to crystals  \cite{zeller1971thermal}. This has been explained phenomenologically invoking the existence of non-linear excitations, dubbed two level systems (TLS) \cite{ anderson1972anomalous}, the nature of which is still elusive \cite{LV13}. 
	A coherent and unifying explanation of the collective emergence of these soft excitations in amorphous solids at low temperature is still lacking. 
	
	Very recently, a new perspective has emerged. The solution of structural glass models in the infinite dimensional limit \cite{SimpleGlasses2020,CKPUZ17} has shown that low temperature glasses may undergo a so called Gardner transition \cite{KPUZ13}, where glassy states become marginally stable. This means that their response to perturbations becomes anomalous and driven by an abundance of soft modes. This approach has been successful to compute the values of the critical exponents of the jamming transition in hard spheres glasses \cite{CKPUZ14} and beyond \cite{FSU19, FSU20, SU21}.
	The Gardner transition therefore may provide the missing ingredient to understand the emergence of soft excitations in low temperature glasses from a first principle perspective.
	
	While for colloidal glasses and relatives, signatures of the Gardner transition have been found in three dimensions \cite{berthier2016growing, seguin2016experimental, JY17, seoane2018spin, jin2018stability, hammond2020experimental, dennis2020jamming,liao2019hierarchical, artiaco2020exploratory}, the Gardner transition in molecular glasses ---described by soft interaction potential between degrees of freedom--- is more elusive \cite{scalliet2017absence, albert2021searching}. 
	Indeed, at the mean field level, the emergence of a finite temperature Gardner transition, upon cooling a glass from high temperature, is accompanied by diverging spin-glass-like susceptibilities, which may be encoded in elastic susceptibilities \cite{biroli2016breakdown} or in non-linear responses to applied electric field \cite{albert2021searching}.
	For molecular glasses, such divergence is not found; something that has been interpreted as a signature of the absence of the Gardner transition \cite{scalliet2017absence} or at least a suppression of Gardner physics \cite{albert2021searching}. 
	
	Finally, at zero temperature, numerical simulations of standard models of molecular glasses have shown a DOS which behaves as $D(\omega)\sim \omega^4$ at low frequencies \cite{Schober_Laird_numerics_PRL, modes_prl_2016, MSI17,richard2020universality,WNGBSF18, kapteijns2018universal,bonfanti2020universal, das2021universal, paoluzzi2020probing}, in contrast with what is predicted by mean field models of the Gardner phase analyzed so far, where $D(\omega) \sim \omega^2$ \cite{FPUZ15, eric_boson_peak_emt}.
	
	This picture changed recently with the introduction in \cite{PhysRevB.103.174202} of a new mean field model of soft, randomly interacting spins, which exhibits a quartic low frequency tail in the DOS at the zero temperature spin glass transition. Correspondingly, the spin glass susceptibility does not diverge and the low frequency tail of the spectrum is populated by localized excitations \footnote{The spin glass susceptibility in this case is given by $\chi_{SG}=\int_0^\infty \de \l \rho(\l)/\l^2$ being $\rho(\l)$ the density of eigenvalues of the Hessian matrix in the minimum of the energy landscape.}. 
	
	The results of \cite{PhysRevB.103.174202} are limited to the phase transition point. Here we characterize throughly the low-temperature glass phase of the model. 
	We show that right beyond the spin glass transition, the glass phase is marginally stable as the Gardner phase of low temperature infinite dimensional models. Nevertheless it is crucially endowed with two types of soft excitations: one associated to a gapless spectrum in the DOS and the other one to pseudogapped non-linear excitations. 
	
	We clarify the mechanism for the emergence of such excitations from the mean field solution of the model, which allows to describe the glass phase in terms of a population of effective spins, subjected to local random effective potentials, whose statistical properties are self-consistently determined. In particular, the effective potential is a quartic polynomial 
	and therefore depending on the corresponding coefficients, it can have a double well shape or a single well shape. 
	We show that the density of effective spins subjected to a potential, having nearly degenerate double wells, is linearly pseudogapped and therefore these spins generate pseudogapped non-linear excitations. Instead, spins subjected to effective potentials, having nearly quartic shape in their ground state, are the source of soft gapless modes in the DOS.

	Hence, we explicitly show that the glass phase of the model introduced in \cite{PhysRevB.103.174202} at the same time encodes both soft linear excitations appearing in the DOS and pseudogapped non-linear ones having the same nature of TLS, as robustly found in low temperature glasses. Furthermore, our picture provides a clear distinction between these two types of excitations, clarifying sharply their difference.
	Finally, our analysis of the glass/Gardner phase shows that the zero temperature spin glass transition in a field found in \cite{PhysRevB.103.174202} is driven by the appearance of non-linear excitations, suggesting a mechanism for the emergence of such transition in finite dimensions.
	
\begin{figure}[t]
		\centering
		\includegraphics[width=\columnwidth]{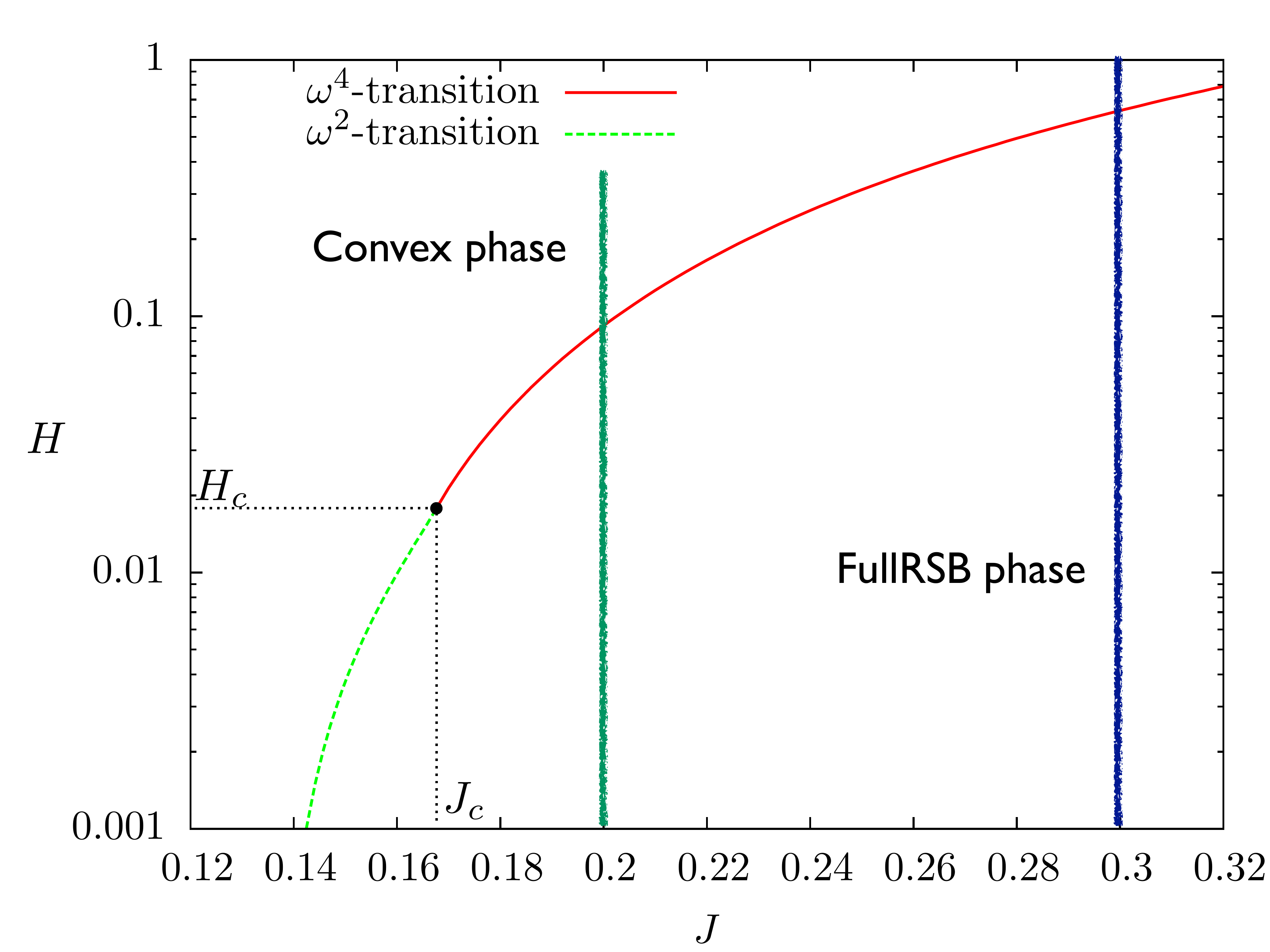}
		\caption{The phase diagram of the model at zero temperature for $k\in[0.1,1]$. Below both the dashed green line and the red continuous line, the model is glassy and described by fullRSB. Above both lines the energy landscape is made by a unique minimum. The properties of the zero temperature spin-glass transition depend on the strength of the external field. For small external field, at the critical point, the model has a quadratic gapless density of states $D(\omega)\sim \omega^2$ while for large field one has $D(\omega)\sim\omega^4$. The two colored cuts are the sections investigated at finite temperature in Fig.\ref{fig:PD}. Data reprinted from \cite{PhysRevB.103.174202}.}
		\label{fig:PD_T0}
	\end{figure}

	\paragraph*{The model and its phase diagram ---} 
	If a Gardner transition arises in molecular glasses, it has been proposed in \cite{albert2021searching} that 
	it could be described as a spin glass phase, emerging from the elastic interactions of groups of degrees of freedom, which can rearrange in different ways (TLS). Therefore following  \cite{GPS03} and \cite{kuhn_and_Horstmann_prl_1997}, we consider the KHGPS model recently introduced in \cite{10.21468/SciPostPhysCore.4.2.008, PhysRevB.103.174202}. The KHGPS model is defined by a set of $N\to \infty$ soft spins $y_i$ taking real values and describing the low temperature effective degrees of freedom as embodied in group of molecules rearranging together in glasses. The spins are subjected to a random local anharmonic quartic potential
	\beq
	v_i(y_i) = \frac 12 k_i y_i^2+ \frac 1{4!}y_i^4 - Hy_i\:.
	\eeq
	The stiffnesses $k_i$ are \emph{i.i.d} random variables extracted from a distribution $p(k)$. Here we will follow \cite{PhysRevB.103.174202} and consider $p(k)$ being a uniform distribution in the interval $[k_{\min}>0, k_{\max}]$ \footnote{Note that the precise shape of $p(k)$ changes our results only qualitative as far as $k_{\rm min}>0$ and $0<p(k_{\rm min})<\infty$.}. The soft spins interact in a mean field manner with all-to-all random couplings, which model the mean field limit of elastic interactions between the effective degrees of freedom emerging at low temperature in molecular glasses. The total Hamiltonian is given by
	\beq
	{\cal H} = -\frac J{\sqrt N}\sum_{i<j} J_{ij}y_iy_j + \sum_{i=1}^N v_{i}(y_i)\:.
	\eeq
	The couplings $J_{ij}$ are Gaussian random variables with zero mean and unit variance.
	The relevant control parameters are the strength of interactions $J$ and the magnetic field $H$. The latter has been introduced to break explicitly the $\mathbb Z_2$ symmetry that plays no role in glasses \cite{BU15, albert2021searching}. The zero temperature phase diagram of the model has been obtained in \cite{PhysRevB.103.174202} and it is reported in Fig.\ref{fig:PD_T0}.

	At fixed $H$, by increasing $J$ the model undergoes a glass transition at zero temperature. For small $J$, the Hamiltonian has a unique minimum while at large $J$, in the glass phase, multiple minima arise. The nature of the zero temperature transition depends on the value of the external field $H$. If $H<H_c$ being $H_c$ a critical value, the glass transition is characterized by the appearance of a gapless spectrum of harmonic excitations, whose low frequency tail behaves as $D(\omega)\sim \omega^2$ and whose eigenmodes are completely delocalized as it happens in previously considered systems \cite{FPUZ15}. For $H>H_c$, the transition point changes nature and it is characterized by a gapless spectrum whose low frequency tail scales as $\omega^4$ and is populated by localized modes. This implies that the nature of the two transitions is different since for $H<H_c$ the transition is accompanied by a diverging spin glass susceptibility, while for $H>H_c$ the zero temperature limit of the spin glass susceptibility is finite.
	In Fig.\ref{fig:PD_T0} we report the phase diagram at finite temperature in the $T-H$ plane at fixed $J$. In the appendix we describe how to get the transition lines and we show that right below the critical temperature, the model exhibits a spin glass phase described by fullRSB, which consequently turns out to be critical.  If $J$ is high enough, the zero temperature transition point at $(0,H_*)$ is in the $\omega^4$ universality class and for the rest of the paper we will focus on this case.

	\begin{figure}[t]
		\centering
		\includegraphics[width=\columnwidth]{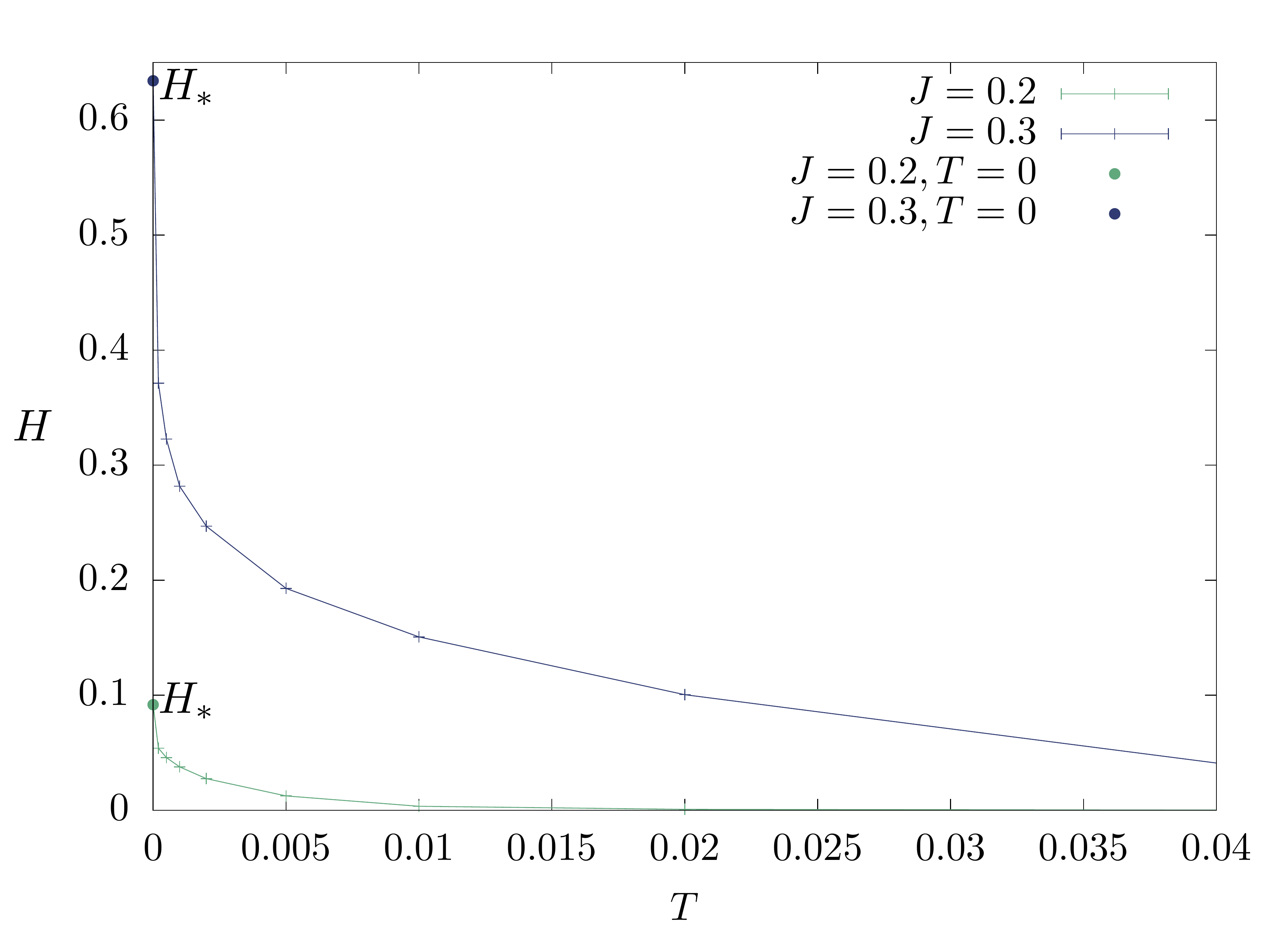}
		\caption{The phase diagram of the model at positive temperature for two values of $J$. 
			As in Fig.\ref{fig:PD_T0}, we fixed $k_{\min}=0.1$ and $k_{\max}=1$. Above the lines the model is described by a replica symmetric (paramagnetic) solution and right below them the solution is fullRSB describing a critical spin glass phase. For both values of $J$, the point at $T=0,\, H=H_*$ is in the $D(\omega)\sim \omega^4$ universality class.}
		\label{fig:PD}
	\end{figure}

	\paragraph*{The replica approach ---}
	The solution of the model can be obtained through the replica method. It boils down to the characterization of the properties of an effective population of spins. Each effective spin $y$ is extracted from a Boltzmann distribution
	\beq
	P(y|h,k)\sim e^{-\beta v_{\rm eff}(y|h,k)}
	\label{eff_v_measure}
	\eeq
	characterized by an effective random potential given by
	\beq
	\begin{split}
		v_{\rm eff}(y|h,k)&=\frac{y^4}{4!} + \frac 12 \tilde k y^2 -hy\\
		\tilde k &= k - \beta J^2(q_d-q)
	\end{split}
	\label{veff_expression}
	\eeq
	being $\tilde k$ the effective stiffness and $h$ the cavity field.
	The replica symmetric solution reported in \cite{PhysRevB.103.174202} gives a way to compute the parameters $q_d,\  q$ and the distribution of the cavity field $h$.
	In particular one has
	\beq
	\begin{split}
		q_d &=\overline{\langle y^2\rangle}^{(h,k)} \ \ \ \ q=\overline{\langle y\rangle^2}^{(h,k)} \\
		P(h)&=e^{-(h-H)^2/(2J^2 q)}/\sqrt{2\pi J^2q}
	\end{split}
	\label{RS_sol}
	\eeq
	and we have indicated with brackets the average over the effective Boltzmann measure of Eq.~\eqref{eff_v_measure}, while the overline indicates the average over $k$, extracted from $p(k)$ and over $h$. Eqs.~\eqref{eff_v_measure},\eqref{veff_expression} and \eqref{RS_sol} provide the replica symmetric solution. 
	The solution is stable as soon as the replicon eigenvalue condition \cite{MPV87} 
	\beq
	\lambda_R=1-J^2\overline{ \left[\frac{\partial^2 f(h|k)}{\partial h^2}\right]^2}^{(h,k)}\geq 0\:.
	\label{rep_RS}
	\eeq
	is satisfied, being $f$ the free energy of the effective spin
	\beq
	f(h|k)= \frac 1\beta \ln \int \de y e^{-\beta v_{\rm eff}(y|h,k)}\:.
	\label{free}
	\eeq
	When the $\omega^4$ transition line is crossed, see Fig.\ref{fig:PD_T0}, the effective stiffness $\tilde k$ can become negative in the zero temperature limit since $J^2\chi=\lim_{\beta\to \infty}J^2\beta(q_d-q)>k_{\rm min}$.
	So, as soon as one extracts a $k<J^2\chi$ so that $\tilde k<0$, the effective potential develops a double well shape if $h$ is close to zero, and therefore in the zero temperature limit Eq.~\eqref{rep_RS} cannot be satisfied since the r.h.s. is negatively divergent.
	
	The glassy phase can be described by a fullRSB solution which can be constructed using the replica method as in \cite{MPV87}.
	The solution boils down to extend the order parameter space from $q_d$ and $q$ to $q_d$ (with the same meaning) and a function $q(x)$ defined in the interval $x\in [0,1]$. The effective stiffness $\tilde k$ is then given by $\tilde k=k-J^2\beta(q_d-q(1))$ and we redefine $\chi=\beta(q_d-q(1))$.
	Furthermore, the distribution of the cavity fields $h$ is now $k$ dependent and given by a distribution $P(x=1,h|k)$, which can be obtained solving the Fokker-Planck equation
	\beq
	\begin{split}
		\dot P &= \frac{J^2\dot q(x)}{2}\left[P''-2\beta x \left(P m \right)'\right]\\
		P(x=0,h|k) &= \frac{1}{\sqrt{2\pi J^2q(0)}} \exp\left[-\frac{(h-H)^2}{2J^2q(0)}\right]\:.
	\end{split}
	\eeq
	where we have denoted with a dot the derivative w.r.t. $x$ and with a prime the one w.r.t. $h$. 
	The function $m(x,h|k)$ obeys the following Hamilton-Jacobi-Bellman equation
	\beq
	\begin{split}
		& \dot m = -\frac{J^2\dot q(x)}{2}\left[m''+\beta x\, m \, m'\ \right]\\
		& m(x=1,h|k) = \frac 1\beta \partial_h \log\int_{-\infty}^\infty\de y e^{-\beta v_{\rm eff}(y|h,k)}
	\end{split}
	\label{Bellman}
	\eeq
	and we note that its boundary condition coincides with the derivative w.r.t. $h$ of Eq.~\eqref{free}. We note also that $m(x=1,h|k)$ is nothing but the magnetization of the effective spin when extracted from an effective potential, having cavity field $h$ and effective stiffness $\tilde k=k-J^2\chi$.
	The solution of these equations is marginally stable in the sense that the corresponding replicon criterion is verified
	\beq
	\lambda_R=1-J^2\overline{\int \de h P(1,h|k) \left[m'(1,h|k)\right]^2}^{(k)}=0\:.
	\label{repRSB}
	\eeq
	We now show how the fullRSB solution cures the instability found at the RS level.
	
	\paragraph*{Zero temperature fullRSB solution ---}
	We will assume that $\chi(H, J)$ is a smooth function across the phase transition which can be proved a posteriori, see the SM. At low temperature, in the broken phase, as soon as $k_{\min}<J^2\chi$ the condition  of Eq.~\eqref{repRSB} cannot be satisfied because of the same mechanism appearing in the RS case. However, if $P(1,h|k<J^2\chi)$ vanishes at $h=0$, the divergence is killed and one may get a finite contribution.
	To clarify this point, we note that for $\beta\gg 1$
	\beq
	m(1,h|k) \sim \begin{cases}
		y_k^*(|h|){\rm sgn}(h) & |h|\gg T\\
		y_k^*(0)\tanh(\beta h y_k^*(0))) & h\sim T
	\end{cases}
	\eeq
	where we have denoted by $y_k^*(h)$ the ground state of the effective potential for a given $h\geq 0$.
	We now assume that for $T\to 0$ and $h\sim T$, $P(1,h|k<J^2\chi)  \sim \gamma_k |h|$. We get that { \medmuskip=-1mu
		\thinmuskip=-1mu
		\thickmuskip=-1mu
		\beq
		\begin{split}
			0&=1-J^2\mathbb E_k\int_{-\infty}^\infty\de h P(1,h|k) \left[\frac{1}{v''_{\rm eff}(y^*_k(|h|)}\right]^2 \\
			&+  J^2\mathbb E_{k<J^2\chi}\int_{-\infty}^\infty \de z (y_k^*(0))^2(1-\tanh^2(z))^2 \tilde P(1,z|k)
		\end{split}
		\label{replicon_full}
		\eeq } being $\tilde P(1,z|k) =\beta P(1,h=zT|k)$. Therefore, the linear pseudogap suppresses the negative divergence responsible for the instability mechanism in Eq.~\eqref{rep_RS}. We observe that (i) no other pseudogap exponent other than linear is as effective, and (ii) if the solution is not of fullRSB type, but it is of a finite RSB type \cite{MPV87} at $T=0$ one can show that $P(1,h=0|k<J^2\chi)>0$ which is unstable because of the negative divergence in the replicon criterion of Eq.~\eqref{repRSB}. So the low temperature solution must be of fullRSB type.
	
	Instead for all $k>J^2\chi$ there is no need to have a pseudogap in $P(x\to 1,h|k)$. 
	The overall picture is that marginal stability is achieved by suppressing the density of states of \emph{resonating} double wells, namely double wells with degenerate minima, whose density is controlled by the cavity fields distribution $P(1,h|k)$. Therefore, in the glassy phase the system is composed by a set of spins for which $k<J^2\chi$ which must be in the ground state of their effective local potential and have non-resonating double wells, and a set of spins for which $k>J^2\chi$ which have a convex effective local potential with a unique minimum. For $k<J^2\chi$ the picture is similar to the Sherrington-Kirkpatrick (SK) model, see \cite{SD84,pankov2006low,crisanti2002analysis}, with the variance that here the emergence of the pseudogap in $P(1,h|k)$ is not induced by hard spins but it is generated by interactions of soft spins through the stiffness scale  $J^2\chi$ which is determined by \footnote{The existence of such a scale was already argued to be essential for quasi localized harmonic excitations through a scaling argument in \cite{10.21468/SciPostPhysCore.4.2.008} and through the RS theory in \cite{PhysRevB.103.174202} where it was identified as the scale responsible for the transition to the glassy phase. Here we are able to give a precise expression for it and show that it crucially controls pseudogapped \emph{non-linear excitations} given by flipping double wells.}
	\beq
	\chi = \overline{\int_{-\infty}^\infty \de h P(1,h|k)m'(1,h|k)}^{(k)}\:.
	\eeq
	We can relate $P(1,h|k)$ with the distribution of the local fields \cite{SD84, thomsen1986local} $h_{\rm loc}^i = \sum_{j\neq i} J_{ij}x_j + H$. If we condition on the local stiffness $k$, $h_{\rm loc}$ is distributed as $h_{\rm loc}\simeq \textrm{sgn}(h)J^2\chi y^*_k(|h|) + h$ being $h$ a random variable extracted from the measure $P(1,h|k)$ \footnote{This is different from the SK model where the Onsager term $J^2\chi y^*(0)$ vanishes at zero temperature.}. While for $k>J^2\chi$ the density of $h_{\rm loc}$ is featureless, for $k<J^2\chi$ 
	the situation changes and the distribution of local fields 
	has a hole around $h_{\textrm loc}=0$ and two linear pseudogaps right outside \footnote{{More precisely, the distribution vanishes in the interval $[-J^2\chi y^*_k(0),J^2\chi y^*_k(0)]$ and it has a linear shape right outside this interval.}}.
	Since Eq.~\eqref{replicon_full} gets a finite contribution from double wells (using the same random matrix arguments of \cite{PhysRevB.103.174202}, see the SM), as soon as we cross the transition point, the spectrum of harmonic excitations is inherited from the distribution of $\tilde a\equiv k-J^2\chi + (y^*_k(h))^2/2$, which is the curvature of the minimum of the effective potential. This is gapless as soon as $J^2\chi\in[k_{\min},k_{\max}]$. If we assume that $P(1,0|k=J^2\chi)>0$, then the ground state right beyond the transition point has a DOS $D(\omega)\sim \omega^4$. The same is found in numerical simulations \cite{10.21468/SciPostPhysCore.4.2.008}.
	In the supplementary material we employ dynamical mean field theory to show that these results mirror what is qualitatively seen in off-equilibrium gradient descent dynamics.
	
	\paragraph*{Discussion ---}
	We have shown that the zero temperature glassy phase of the KHGPS model right beyond the glass transition point is marginally stable ---as the Gardner phase of infinite dimensional models of glasses--- and in addition, it is described by a mixture of gapless linear and pseudogapped non-linear excitations.  We believe that the relevant question now is the characterization of the zero temperature transition point.
	Indeed, the appearance of a spin glass transition in a field in finite dimension is a debated issue of huge relevance, and given that the Gardner transition has precisely the same nature \cite{BU15}, establishing its existence may have a big impact in our understanding of low temperature glasses. 
	It is believed that if a transition arises in finite dimensions, it must be described, in a renormalization group sense, by an attractive fixed point at zero temperature \cite{parisi2012replica}. Our analysis suggests that the mechanism behind the emergence of this transition is the appearance of non-linear excitations, as we have found in our model.
	
	We underline that the formal properties of the zero temperature critical point are very interesting. 
	It is believed that one of the crucial quantities that describes the corresponding critical theory in finite dimension is the so-called $\lambda$-parameter  \cite{parisi2014diluted}. In our model, this parameter is finite at the zero temperature critical point (see the SM). Instead, the SK model which lacks of a critical point at finite external field, has precisely the same property \cite{temesvari1989thermodynamics,crisanti2003parisi} if the field is allowed to diverge.
	In order to make further progress, an essential step now is to develop a renormalization group computation for this critical point \cite{PhysRevB.65.224420}. To do so, one needs to control the diverging susceptibilities. At zero temperature, thermal fluctuations are absent and one is left with sample-to-sample fluctuations which must be carefully investigated.
	
	\paragraph*{Acknowledgments ---} 
	This  work  is  supported  by \textquotedbl Investissements d'Avenir\textquotedbl \    LabExPALM (ANR-10-LABX-0039-PALM). PU warmly thanks Thibaud Maimbourg for enlightening discussions. The authors thank E. Bouchbinder, E. Lerner, M. M\"uller, C. Rainone, V. Ros, M. Wyart and F. Zamponi for discussions.

	\bibliography{refs_with_doi.bib}

\widetext
\appendix

\section{The finite temperature spin glass transition}
The finite temperature transition line as in Fig.~2 of the main text can be obtained by integrating numerically Eq.~(5) of the main text and checking the stability of the replicon condition (Eq.~(6) of the main text), which signals the spin glass transition. To characterize the nature of the low temperature phase one needs to look perturbatively at the properties of $q(x)$ right beyond the transition point. In this case, the solution for $q(x)$ deviates from a simple constant. In particular one can define the so called breakpoint of $q(x)$ as the point $x=x^*$ at which $q(x)$ deviates from a flat constant, see \cite{SimpleGlasses2020}. Furthermore, since $q(x)$ deviates from a constant at $x=x^*$, one have $\dot q(x^*)\neq 0$. The low temperature phase is of fullRSB type as soon as $\dot q(x^*)>0$ and $x^*\in [0,1]$. \cite{sommers1985parisi, SimpleGlasses2020}. The breakpoint $x^*$ is also called the $\lambda$-parameter of the transition \cite{parisi2014diluted}.
One can give a closed expression for both $x^*$ and $\dot q(x^*)$ at the transition point \cite{sommers1985parisi, SimpleGlasses2020}.
We have

\begin{equation}
x^* = \frac{\overline{\int_{-\infty}^{\infty}dh P(0,h|k)m''(1,h|k)^2}^{(k)}}{2\overline{\int_{-\infty}^{\infty}dh P(0,h|k)m'(1,h|k)^3}^{(k)}} = \frac{\overline{\mathbb{E}^{RS}_{h} K_3(y)^2}^{(k)}}{2\overline{ \mathbb{E}^{RS}_{h} K_2(y)^3}^{(k)}}
\end{equation}
\begin{equation}
\begin{split}
\dot{q}(x^*) &=\frac{2\overline{\int_{-\infty}^{\infty}dh P(0,h|k)m'^3}^{(k)}}{\overline{\int_{-\infty}^{\infty}dh P(0,h|k)\big ( {m'''}^2-12x^*m'm''^2+6{x^*}^2{m'}^4\big )}^{(k)}} \\
\end{split}
\end{equation}
where
$$\mathbb{E}^{RS}_{h}  A =\int_{-\infty}^{\infty}dh P(0,h|k) A(h)$$
and
\begin{equation*}
\begin{aligned}
K_1(y) &= \langle y\rangle\\
K_2(y) &= \langle y^2 \rangle -\langle y\rangle^2\\
K_3(y) &= \langle y^3 \rangle -3\langle y^2\rangle\langle y\rangle +2\langle y\rangle^3\\
K_4(y) &= \langle y^4 \rangle -4\langle y^3\rangle\langle y\rangle -3\langle y^2\rangle^2+12\langle y^2\rangle\langle y\rangle^2-6\langle y\rangle^4\\	
\end{aligned}
\end{equation*}
where
$$\langle A(y) \rangle=\frac{\int_{-\infty}^{\infty}dy e^{-\beta v_{\text{eff}}(y|h,k) }A(y)}{\int_{-\infty}^{\infty}dy e^{-\beta v_{\text{eff}}(y|h,k) }}$$

\begin{figure}[h]
	\centering
	\includegraphics[width=0.45\textwidth]{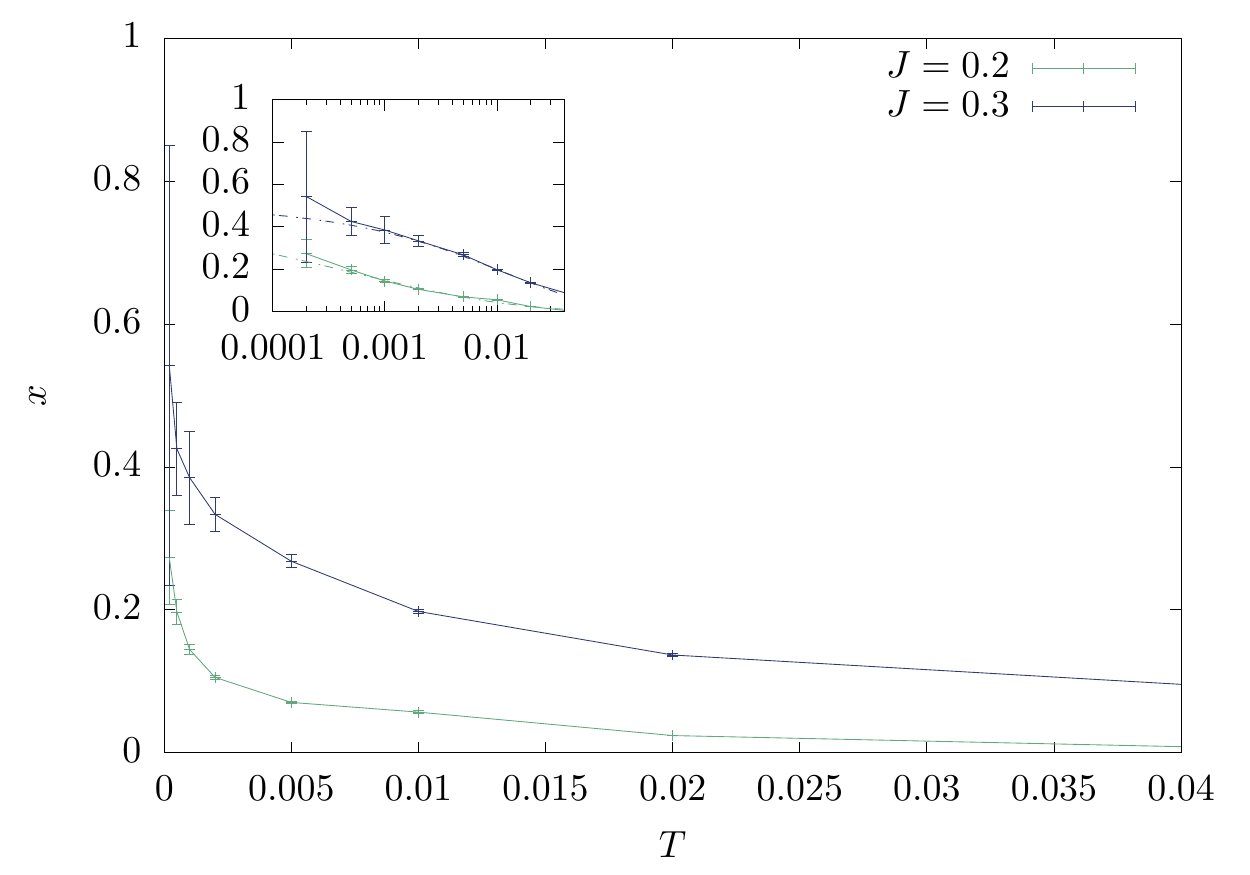}
	\includegraphics[width=0.45\textwidth]{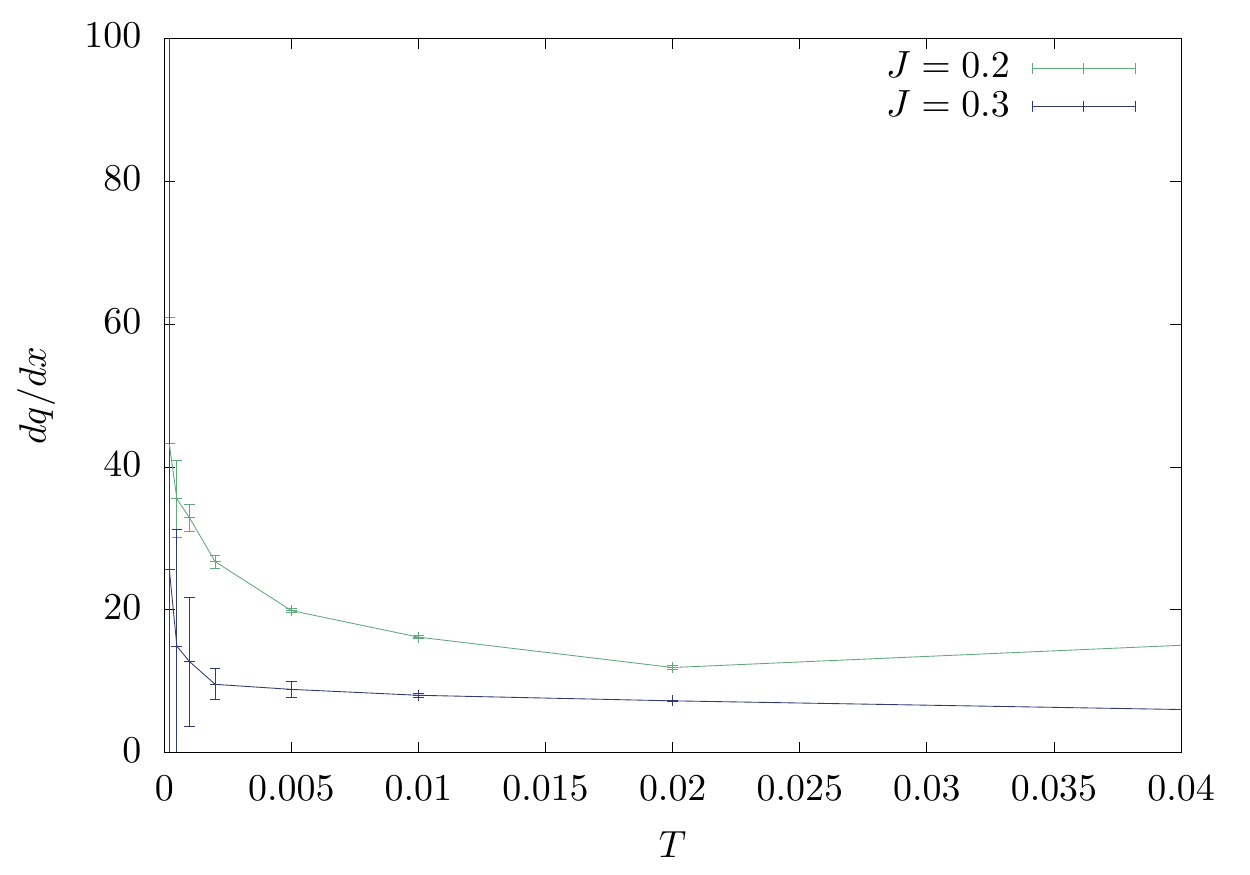}
	\caption{We computed numerically the breakpoint $x^*$ and slope of $q(x^*)$ at the transition point as a function of the temperature. Both these quantities show that the transition is of continuous fullRSB type. Remarkably, at zero temperature the breakpoint $x^*$ tends to a finite value. The error bars are due to Monte Carlo evaluation of the integrals that define these quantities. In the inset of the right panel we show a tentative fit with the function $0.5e^{-A T^\nu}$ compatible with $x^*=1/2$ at the transition. A more careful analysis is needed to compute precisely the value of the breakpoint.}
	\label{fig:x_qdot}
\end{figure}

In Fig.\ref{fig:x_qdot} we plot the behavior of $x^*$ and $\dot q(x^*)$ at the transition point as a function of the temperature.
We observe that for the two values of $J$ shown in Fig.2 of the main text (vertical lines), the perturbative solution gives a fullRSB phase beyond the transition.
Furthermore, we underline that in the zero temperature limit, when $H\to H_*$ one gets that $x^*\to \sim 0.5>0$. This is very different from \cite{FPSUZ17} or the KHGPS model at small external field (below $H_c$). Interestingly it is close to what happens in the SK model on the critical line, for vanishing temperature, where the breakpoint is equal to $1/2$ as shown in \cite{temesvari1989thermodynamics,crisanti2003parisi}.
Finally we notice that, at any finite temperature, the spin glass susceptibility is divergent at the transition point. Close to the transition one has:
\beq
\chi_{SG} \sim \frac{A(T,H)}{|T-T_c(H)|}
\eeq
When the zero temperature critical point ($H=H_*$) is in the $D(\omega)\sim \omega^4$ universality class, one has that $T_c(H_*)=0$ and $A(T,H_*)\sim T$, and the spin glass susceptibility does not diverge anymore. This implies that the spin glass transition at zero temperature is of very different nature.

\section{Tail of the harmonic spectrum}
The spectrum of the model is encoded in the Hessian matrix as defined as
\beq
M_{ij}=J_{ij} + v_i''(y_i)\delta_{ij}
\label{hessian}
\eeq
being $y_i$ the ground state of the model.
The tail in the harmonic density of states can be understood as follows. Right beyond the transition point we have
\beq
1>J^2\mathbb E_k\int_{-\infty}^\infty\de h P(1,h|k) \left[\frac{1}{v''_{\rm eff}(y^*_k(|h|)}\right]^2\:.
\label{LAMBDA}
\eeq
Furthermore we can easily show that the scaling region of $h\sim T$ in the double wells range $k<J^2\chi$ does not give contribution to the equation for $\chi$ which can be written as
\beq
\chi = \mathbb E_k\int_{-\infty}^\infty\de h P(1,h|k) \frac{1}{v''_{\rm eff}(y^*_k(|h|)}
\label{chi_RS}
\eeq
Eqs.~\eqref{LAMBDA} and \eqref{chi_RS} imply that the spectrum is dominated by the random diagonal matrix in Eq.~\eqref{hessian} (as in \cite{PhysRevB.103.174202}) meaning that it is gapless if $J^2\chi\in[k_{\min}, k_{\max}]$ and its tail is inherited from the distribution of effective masses $\tilde a=k-J^2\chi+(y^*)^2/2$ which are therefore induced by the cavity field distribution.

\section{Dynamical mean field theory}
\begin{figure}[t]
	\centering
	\includegraphics[width=0.45\textwidth]{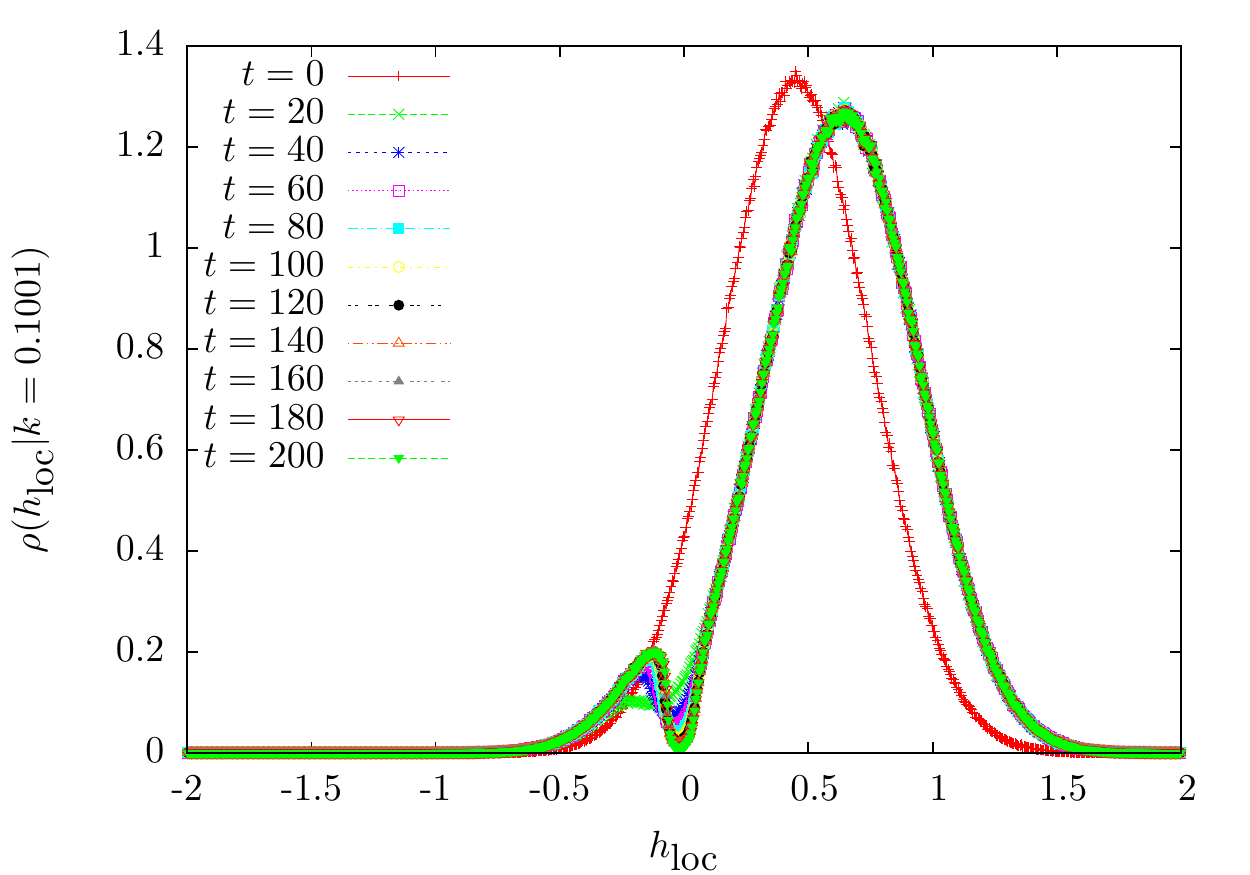}
	\includegraphics[width=0.45\textwidth]{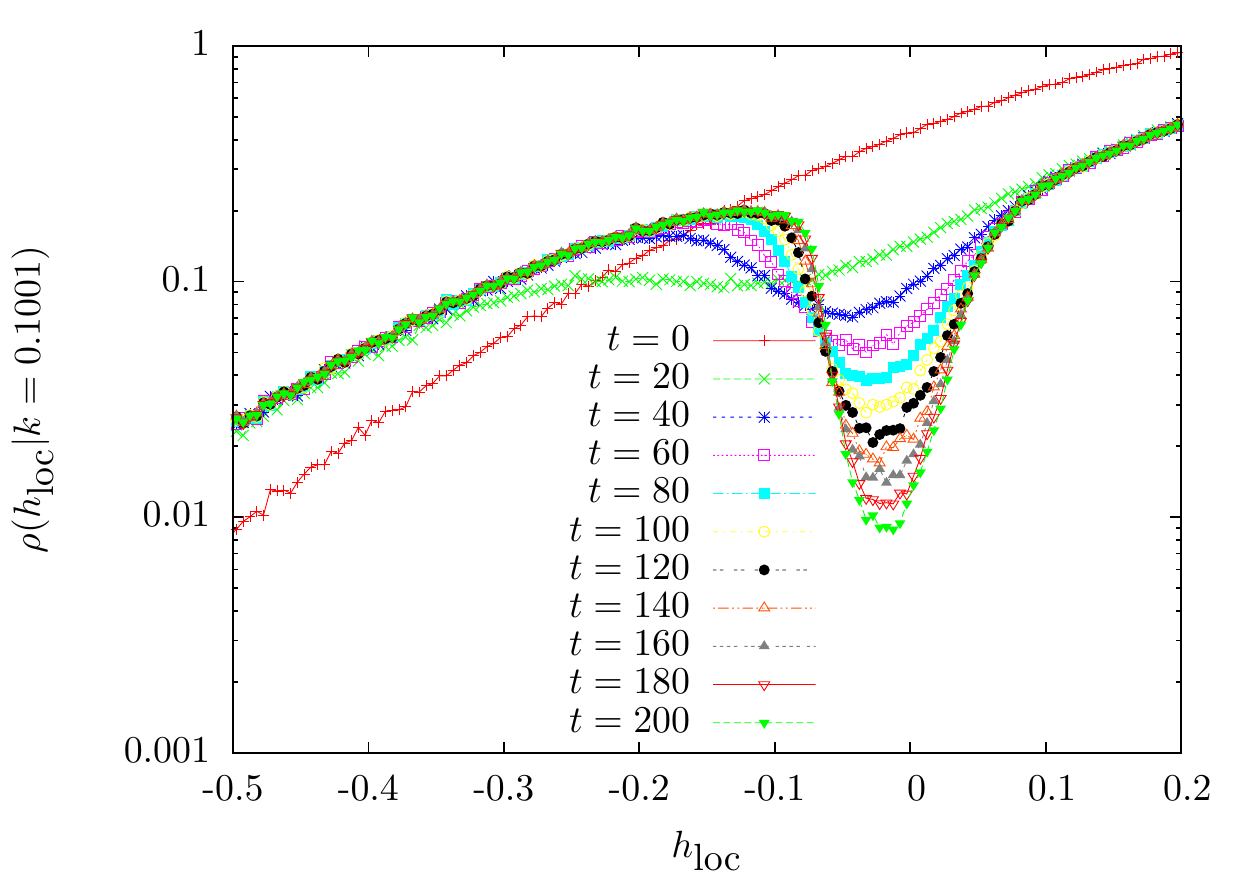}
	\includegraphics[width=0.45\textwidth]{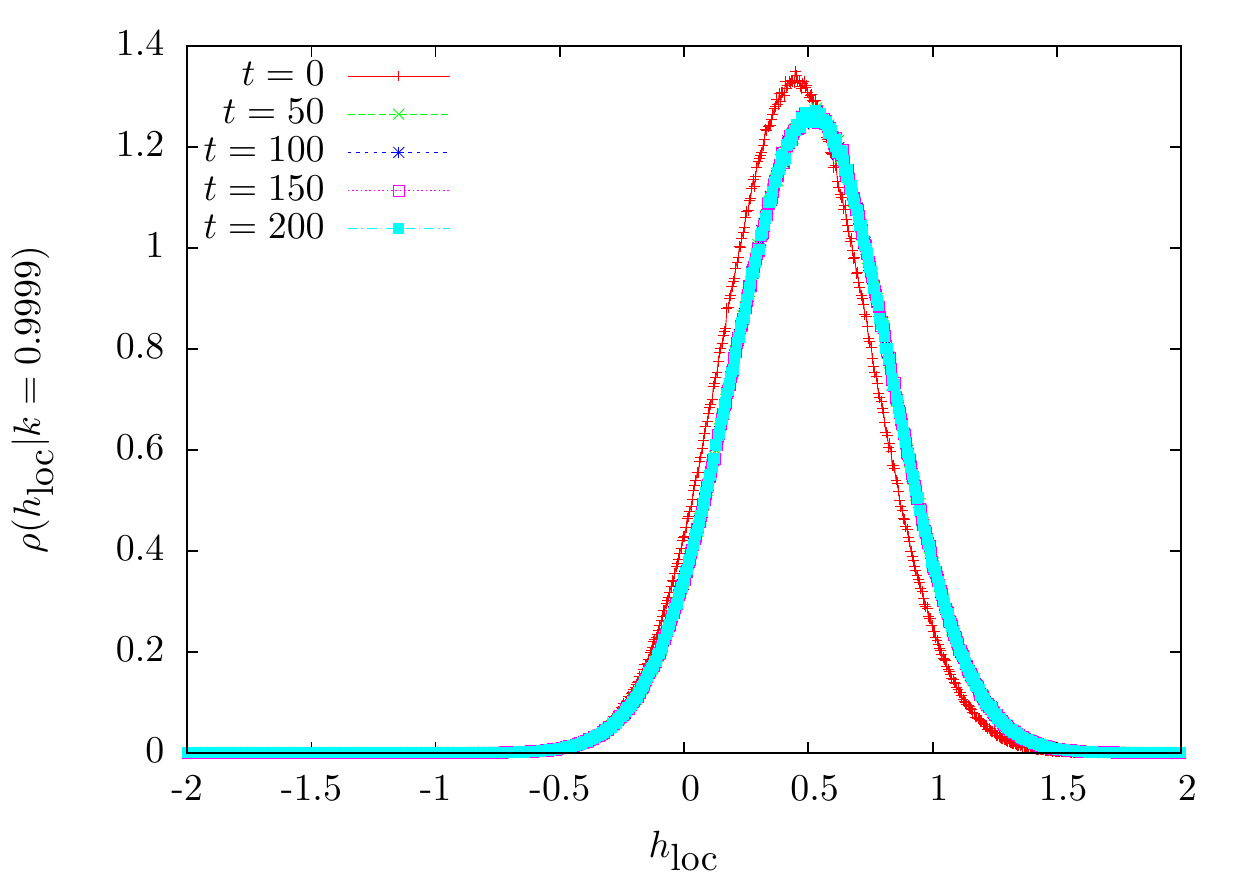}
	\caption{The distribution of local fields computed by the numerical integration of the DMFT equations for $J=0.3$ and $H=0.45$. For $k=0.1001$ (top panels) a depletion is opening around $h_{\textrm{loc}}=0$ when time increases. The panel on the left shows the full distribution, while the panel on the right shows a zoom on the hole which opens as time increases. Conversely for $k=0.9999$ (bottom panel) the distribution is regular and converges to a Gaussian-like shape. We emphasize that, while for large $k$ the local field distribution converges very fast, for small $k$, on the timescales we have access to, the dynamics is far from asymptotic and the hole is still not well formed.  }
	\label{fig:local_fields}
\end{figure}

\begin{figure}[t]
	\centering
	\includegraphics[width=0.5\textwidth]{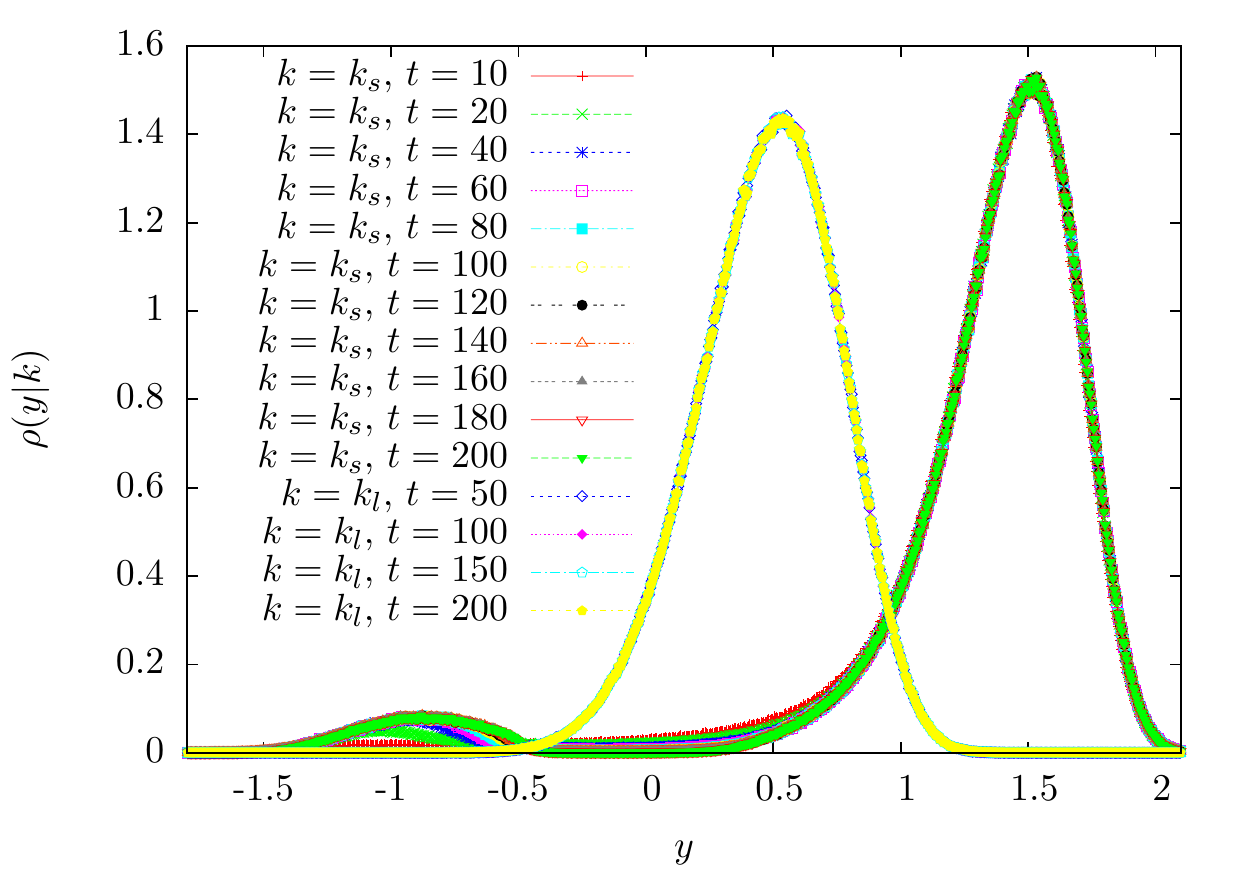}
	\caption{The distribution of spins. For $k=k_s$ the distribution has a hole opening around $y=0$ and a double peak structure. For large $k=k_l$ the distribution is featureless.}
	\label{fig:positions}
\end{figure}

To explore numerically the glassy phase we use gradient descent dynamics. We initialize all spins at $y_i(t=0)=1$ \footnote{Any other random initial condition, separable on the degrees of freedom, does not affect the final result.} and run a gradient descent dynamics as
$\dot y_i(t) = -{\partial \cal H}/{\partial y_i}$. For $N\to \infty$, the dynamics can be analyzed 
through dynamical mean field theory (DMFT). The resulting equations are a straightforward extension of the ones reported in \cite{MPV87}. 
The $N\to \infty$ dynamics is described by a self-consistent stochastic process for an effective spin
\beq
\dot y(t) = -v_k'(y) + \int_0^t\de sM_R(t,s)y(s)+\eta(t)
\label{effective_process}
\eeq
where the noise $\eta(t)$ is centered and Gaussian and has a two point function obtained self consistently as $\langle\eta(t)\eta(t')\rangle = M_C(t,t') = J^2\mathbb E_k \langle y(t) y(t')\rangle$.
The memory kernel instead is given by $M_R(t,t') = J^2\mathbb E_k \left\langle{\delta y(t)}/{\delta\eta(t')}\right\rangle$.  We integrate numerically eq.~\eqref{effective_process} using the same algorithm of \cite{EO92,MKUZ20} thus obtaining $M_C$ and $M_R$. 
We then focus on the local fields distribution defined as $h_{\textrm loc}(t) = \int_0^t \de s M_R(t,s)y(s)  +\eta(t) + H$. One can show that the histogram of $h_{\textrm{loc}}(t)$, $\rho(h_{\textrm{loc}}|k)$ coincides in the $N\to \infty$ limit with the statistics of $h_{\textrm{loc}}^i = J\sum_{j \neq i} J_{ij}y_j(t)/\sqrt N + H$.
The advantage of using DMFT is that the statistics of the local fields can be obtained at fixed $k$, meaning that we sample the local fields conditioning on the stiffness of the corresponding spins. In Fig.~\ref{fig:local_fields} we show the distribution of local fields, for two values of $k$, at different times. At long times, for $k$ slightly larger than $k_{\min}$, we find a depletion of $\rho(h_{\textrm loc}|k)$ around the origin. For $k$ slightly smaller than $k_{\max}$, $\rho(h_{\textrm{loc}}|k)$ converges to a regular shape and it is finite in the origin.
This behavior is mirrored by the statistics of typical spins $y(t)$, $\rho(y|k)$, which has a depletion around the origin and two peaks for small $k$, while it has a single peak for large $k$ (see Fig.\ref{fig:positions}).

\begin{figure}[h]
	\centering
	\includegraphics[width=0.45\textwidth]{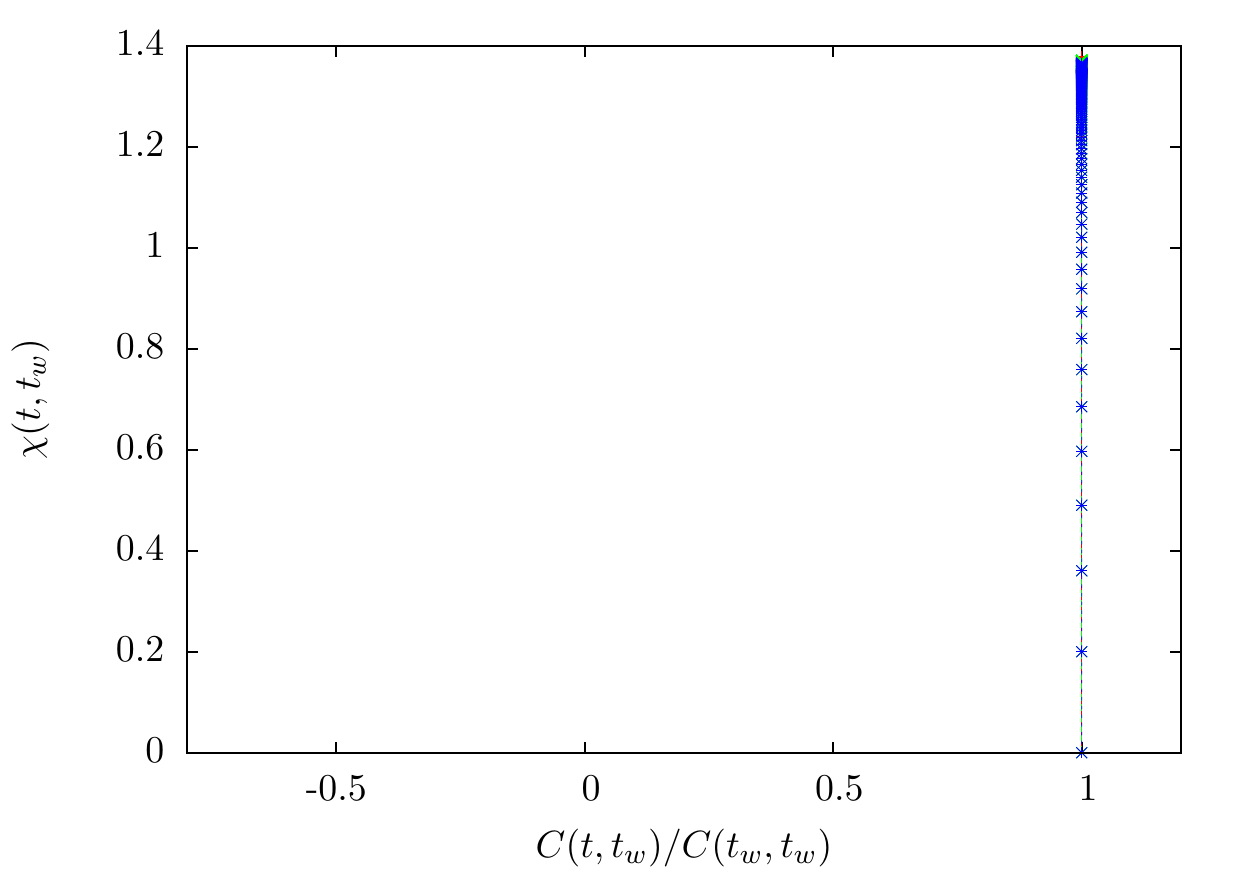}
	\caption{The integrated response for $H=0.45$ and $J=0.3$, plotted against the correlation function for long times and different $t_w=100,140,160$. The curves are all coincident.}
	\label{fig:FDT}
\end{figure}

Finally, we consider the fluctuation dissipation ratio (FDR) as defined from the fluctuation dissipation theorem (FDT) connecting correlation and response \cite{cugliandolo2002dynamics}. In Fig.\ref{fig:FDT} we plot the integrated response $\chi(t,t_w)=\int_{t_w}^t\de s M_R(s,t_w)/J^2$ for different waiting times $t_w$ (up to time $t=200$) as a function of $C(t,t_w)=M_C(t,t_w)/J^2$. The slope of $\chi(t,t_w)$ in this parametric plot defines the FDR \cite{cugliandolo2002dynamics} and we observe that, on the timescales we have access to, this is vertical, implying an infinite FDR.

\end{document}